\documentclass[10pt, conference, compsocconf]{IEEEtran}
\usepackage{cite}

\ifCLASSINFOpdf
   \usepackage[pdftex]{graphicx}
   \graphicspath{{../pdf/}{../jpeg/}}
   \DeclareGraphicsExtensions{.pdf,.jpeg,.png}
\else
  \usepackage[dvips]{graphicx}

   \graphicspath{{../eps/}}
   \DeclareGraphicsExtensions{.eps}
\fi
\usepackage{amsmath}
\usepackage[english]{babel}
\usepackage[utf8]{inputenc}
\usepackage{algorithm}
\usepackage[noend]{algpseudocode}

\usepackage{array}
\begin{document}
%
\title{Machine Identification of High Impact Research through Text and Image Analysis}
%
%
%
\author{\IEEEauthorblockN{Marko Stamenovic}
\IEEEauthorblockA{\textit{University of Rochester}\\
\textit{Rochester, NY}\\
\textit{mstameno@ur.rochester.edu}}
\and
\IEEEauthorblockN{Jiebo Luo}
\IEEEauthorblockA{\textit{University of Rochester}\\
\textit{Rochester, NY}\\
\textit{jluo@cs.rochester.edu}}
}

\markboth{2016 IEEE International Conference on Big Data}%
{Shell \MakeLowercase{\textit{et al.}}: Bare Demo of IEEEtran.cls for IEEE Journals}
%



\maketitle

\begin{abstract}
The volume of academic paper submissions and publications is growing at an ever increasing rate. While this flood of research promises progress in various fields, the sheer volume of output inherently increases the amount of noise. We present a system to automatically separate papers with a high from those with a low likelihood of gaining citations as a means to quickly find high impact, high quality research. Our system uses both a visual classifier, useful for surmising a document's overall appearance, and a text classifier, for making content-informed decisions. Current work in the field focuses on small datasets composed of papers from individual conferences. Attempts to use similar techniques on larger datasets generally only considers excerpts of the documents such as the abstract, potentially throwing away valuable data.  We rectify these issues by providing a dataset composed of PDF documents and citation counts spanning a decade of output within two separate academic domains: computer science and medicine. This new dataset allows us to expand on current work in the field by generalizing across time and academic domain. Moreover, we explore inter-domain prediction models - evaluating a classifier's performance on a domain it was not trained on - to shed further insight on this important problem.  
\end{abstract}

\begin{IEEEkeywords}
Document Classification, Education, Computer Vision, SIFT, SVM, K-Means,  Text Analysis, Data Mining.
\end{IEEEkeywords}

%
\IEEEpeerreviewmaketitle

\section{Introduction}
%
%
%
%
Citation counts have traditionally been used as a proxy for influence, innovation and quality of academic papers\cite {impact}. They are also used, along with productivity, to rate the output of a researcher using the h-index, \cite{h-index3,h-index2} a metric that defines a scientist's impact based on the number of articles they have published and citations they have accrued. According to Hirsch, a scientist publishing 12 papers with at least 12 citations each could qualify for tenure at a major university \cite{h-index}. Although an academic paper which never receives citations may be of intrinsic value,  papers which receive more citations will generally have a more significant effect on the academic and R\&D community at large.

In this paper we present a system which, given a database of academic papers in a specific field and the associated citation counts, attempts to predict whether a future paper will be of high impact based on extracted visual and textual features. We also explore the problem of inter-domain classification, in which a classifier is evaluated on a dataset dissimilar from the content it was trained on, i.e. feeding a classifier trained on medical journal articles with a dataset of computer science journal articles. We find that training on text features generally provides significantly higher performance than only using visual features. However, while using  text features  provides a stronger signal for intra-domain classification, visual features improve the robustness of the classifier when moving across domains, indicating that visual features may provide a more universal overall signal for this classification task.

One application of such a system includes a smart search engine for quickly identifying high-impact research among recently published papers which have not yet had time to accrue citations. These  papers may be relatively easy to find for researchers in the field or with high domain-specific knowledge but they are elusive to those who are new to a field, or those who are doing cross-disciplinary research in an unfamiliar one.  Such a system could also be valuable to government and industry in order to quickly aggregate knowledge about the direction of research in order to make better informed policy or business decisions.

The remainder of the paper is laid out as follows. In Section 2 we discuss some of the related work in the field. In Section 3 we review publicly available datasets, explain why they are not useful for our method and describe how we constructed our own. In Section 4 we describe how we extracted text and visual features from the dataset. In Section 5 we describe how we built our classifiers. In Section 6 we present our experimental results. In Section 7 we present a discussion of our work, observations and potential drawbacks. Finally, in Section 8 we present our conclusions. 
\begin{figure*}
  \label{fig:kpts}
  \includegraphics[width=\textwidth,height=6cm]{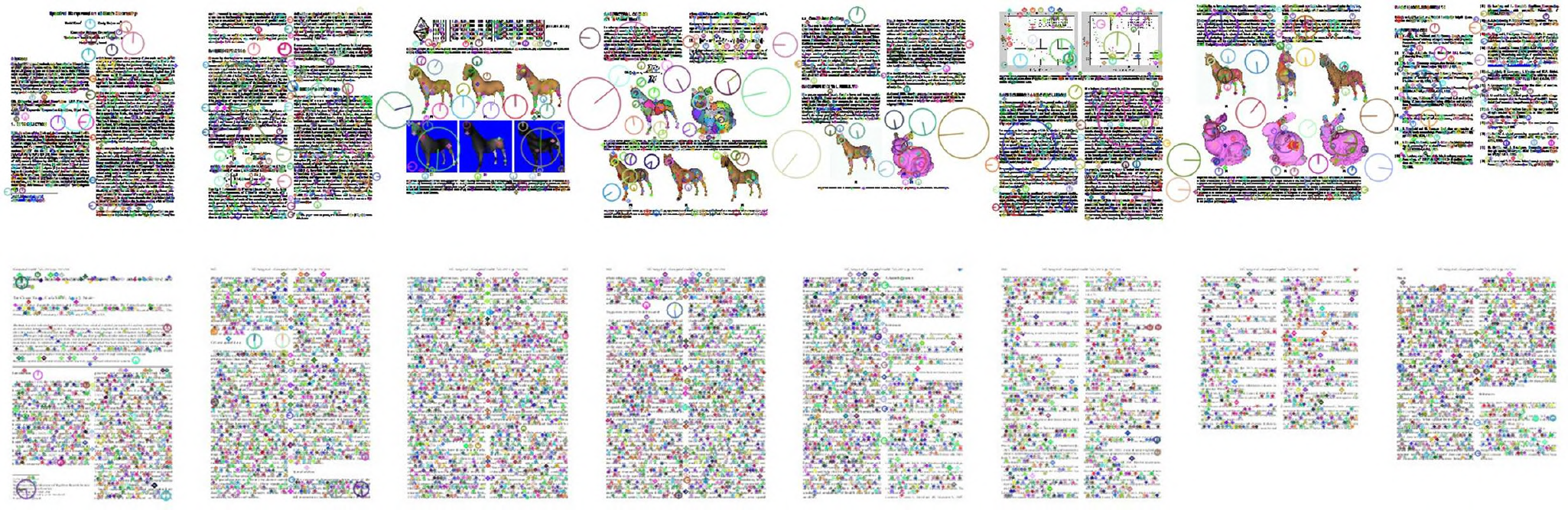}
  \caption{A visualization of the keypoints extracted by the SIFT algorithm. Keypoints are indicated by the different colored and sized circles overlayed on the papers. The top paper is a a correctly classified high citation paper while the lower paper is a correctly classified low citation paper. By examining the figure it is visible that many of the keypoints in the high citation paper lie in the region of the images indicating that the algorithm is learning to classify the images as a feature of high citation papers. In addition,  keypoints identified in both papers include text, equations, code and white-space.}
\end{figure*}

\section{Related Work}
The literature with respect to academic document classification and knowledge extraction is rich. In \cite{impact}, McNamara et al. seek to predict high-impact papers using citation network analysis. This method differs from ours in that it uses author network analysis to predict a paper's impact rather than the content of the paper itself. In \cite{nature} Acuna et al. seek to predict a scientist's future h-index using linear regression on past citation data with favorable results. In \cite{heirarchicalclassifiers}, Ruiz and Srinivasan present a system to classify the category of a medical paper based on text features extracted from its abstract by combining a linear classifier with a simple neural network. In \cite{siftretrieval}, Smith and Harvey use Scale Invariant Feature Transform (SIFT) visual features academic papers to classify them by category. In \cite{gestalt}, the authors have success in predicting whether a paper is a workshop paper or an accepted conference paper using a support vector machine (SVM) classifier on clustered SIFT features from the the first seven pages of the paper, thus capturing the gestalt of the paper. For clarity, ``gestalt" in our usage is defined as the overall perceptual structure of an academic paper; not only the content but also the space separating it and the presentation\cite{definegestalt}.

    Our method improves and expands on the work above in the following ways: First, our system uses both text and visual features extracted from the entirety of the document as opposed to\cite{gestalt}, where the authors normalize the paper length to seven pages by either truncating or adding white space. Truncating a document can deeply alter a its gestalt; for example, longer papers will likely have their conclusions and results tables omitted. Additionally, only using text from the abstract, as in \cite{heirarchicalclassifiers}, drastically reduces the amount of features available to classify each paper. Our method, which uses text features from the entire paper affords a much richer feature set. It also uses a meta-classifier to fuse together the visual classifier with the text classifier, which improves the overall performance while adding a layer of robustness to the results, especially across domains. We use paper content in addition to citation count to identify high quality research, whereas \cite{nature,networks} only use citation count. We also expand on \cite{gestalt} and \cite{siftretrieval} by increasing the breadth and scale of the dataset to cover citation counts in entire fields of study over a 13-year period, rather than accepted vs. workshopped at isolated conferences. Finally, we explore both intra-domain and inter-domain prediction models to shed further insight on the problem of using a classifier to predict outside of the domain it was trained on. 


\IEEEpubidadjcol


\section{Data Sources}
	Publicly available datasets which have been used to study academic document mining in the past include OHSUMED\cite{OHSUMED} which itself is a subset of MEDLINE \cite{datasets} and the Citeseer Dataset \cite{citeseerdataset}. These datasets contain paper abstracts, keywords, authors and citation counts. Other academic datasets include the NIPS dataset\cite{nipsdataset}, which contains raw text files of the entire paper for the first 12 years of the NIPS conference. However, none of the publicly available datasets contain the raw document PDF files and citation counts, which are necessary to extract features for our analysis - the document's entire text, an image file of the entire document and citation count.
    
    Therefore, we used CiteSeer$^x$, \cite{citeseer} an online scientific digital library of freely available academic papers to build our own dataset.  CiteSeer$^x$ is an automated system that creates an index of academic articles, complete with an interactive bibliography, citation count, and links to freely-available PDF downloads. A web spider was written to automatically download the article PDF's  along with pertinent metadata such as citation counts, article names and authors in two distinct fields: computer science and medicine. We specified that the spider would only download an article if it had received 0 citations, corresponding to a low-impact paper, or more than 10, corresponding to a high impact paper. The high citation boundary of 10 was chosen according to Guerrero-Bote and Moya-Anegon \cite{citationsvsyears}, who show that an ``excellent'' paper, defined as a paper in the 90th percentile in its field in terms of citations counts, will receive at least 10 citations after its first three years in publication. No articles after 2013 were used in order to allow them a reasonable amount of time to accrue citations. Additionally, a fixed date range was used in order to keep the articles on an even footing in terms of graphic technology and culture. Thus, we limited our search to articles with a publication date from 2000 to 2013. 
    
    Our fields of study, or domains, were defined within the CiteSeer$^x$ framework by keyword queries. For example, keywords for the medical domain included terms such as ``pathology," ``medicine,"  and ``genomics" while keywords for the computer science domain included terms such as ``computer science," ``information security," and ``machine learning." We generated these keywords using the subfields of computer science and medicine according to Microsoft Academic \cite{microsoft}.
    
Our scrape of  CiteSeer$^x$ turned up 1,894 medical articles and 5,785 computer science articles, each balanced to contain approximately 50\% high citation papers and 50\% low. Our dataset increases in size compared to similar studies in \cite{gestalt} and \cite{siftretrieval} by a factor of 4 and 25, respectively. Although scraping articles from CiteSeer$^x$ is likely to provide noisier output than a handmade dataset \cite{citeseer2}, it would be impossible to collect the same amount of information in a reasonable amount of time without automation.
\section{Feature Extraction}
	\subsection{Text Features}
We extracted text data from each PDF file and then tokenized the extracted text by down-casing, removing  punctuation and removing all English stop words (such as ``the" ``or" ``and"). Stop-words are devoid of meaning and thus are useless to our classification task.  We also removed any text indicative of a year by filtering out tokens matching the date range between 1950 and 2017. This was done in order to prevent skewing the classifier based on increased age and therefore increased time to accrue citations. We built word frequency vectors for each PDF based on the described tokenization procedure. The vectors were then weighted by a term frequency-inverse document frequency (TF-IDF)\cite{term-frequency, document-frequency} filter. TF-IDF is a weighting scheme used to give more importance to vectors which occur frequently in a particular document and simultaneously give less weight to vectors which occur across many documents in the corpus. Down-weighting terms which occur more frequently across a corpus has the effect of decreasing the classification contribution of more common, less interesting words, while up-weighting the importance of rarer, more interesting words. TF-IDF is calculated as shown below:

\[tf\mbox{-}idf(t,d) = tf(t,d) \times idf(t) \label{eqn:tfidf}\tag{1}\]
where \(tf(t,d)\) is the term frequency of word \(t\) in document \(d\) and \(idf(t)\) is the inverse document weighting of term \(t\) across the entire corpus. The inverse document frequency \(idf(t)\) is calculated as follows:

\[idf(t) = \log \frac{1 + n_d}{1+df(d,t)}\label{eqn:idf}\tag{2}\]
where \(n_d\) is the total number of documents and \(df(d,t)\) is the number of documents that contain term \(t\). The resulting vectors \(v\) are then normalized by the L2 norm as shown in equation \eqref{eqn:l2reg} to provide a more consistent feature space for the classifier.

\[v_{norm} = \frac{v}{\left \| v \right \|_2} = \frac{v}{\sqrt{v_1^{2}+v_2^{2}+ ... +v_n^{2}}}\label{eqn:l2reg}\tag{3}\]
Finally, the feature set was truncated to the top 50k by term frequency across the corpus as a form of regularization.

\subsection{Visual Features}
The first step in visual feature extraction was pre-processing the raw PDF's. The PDF's were converted to JPEG using a constraint of 300px height. Each page in a document was horizontally concatenated to ensure that each article would be viewed by the classifier as one piece of data.
	
The scale invariant feature transform (SIFT) \cite{sift} algorithm was originally developed to extract image features invariant to scale and rotation for use in matching between different views of an object or scene. We chose to use SIFT features to represent the images due to their demonstrated success in similar classification tasks in the literature \cite{gestalt,siftclass,bof,siftretrieval}. When all SIFT descriptors were extracted and stacked the resulting array had dimensions larger than $5.1\times10$$^6\times128$. In order to make the problem tractable, a more moderately sized visual vocabulary was needed to describe the visual features of the documents.  

Our solution was to use k-means clustering\cite{kmeans} to reduce the dimensionality of the raw SIFT vectors. The k-means algorithm partitions a set of \(i\) samples \(x\) into \(k\) disjoint clusters \(\textbf{C} = \{C_1,C_2...C_k\}\), each described by the mean \(\mu_i\) of samples in the cluster. The means are commonly referred to as cluster "centroids". The goal of k-means is to minimize the within-cluster sum of euclidean distances to the centroid, or in other words to solve the equation in \eqref{eqn:kmeans} where \(\mu_i\) is the mean of points in \(C_i\).
\[\arg \min_\mathbf{C} \sum_{i=1}^{k} \sum_{\mathbf{x} \in C_i} \left \| \mathbf{x} -\mathbf{\mu}_i \right \|_2\label{eqn:kmeans}\tag{4}\]
Raw SIFT features for each document were mapped to the centroid to the nearest cluster and then grouped into a frequency vector based on their number of occurrences in the document. A size of \(k=100\) was chosen for the number of centroids across the corpus by iteratively increasing cluster size by a factor of 3 and evaluating classification results on the development set.  When classifier performance began to decrease with respect to number of centroids, indicating over-fitting, the number of centroids was fixed at the previous value. A visualization of the extracted SIFT features overlaid on a document is shown in Figure 1. The final step in visual feature extraction was standardization of the features in order to prepare the data for ingestion into the classifier using mean-removal and variance-scaling. 
	
\section{Classifier Construction}
    
    \subsection{Text Classifier}
    For text classification, we used a support vector machine (SVM)\cite{svm2} model with a linear kernel on the input text frequency vectors and binary output classes - either high citation or low citation. An SVM constructs support vectors in the high dimensional input space which are then optimized to create an optimal hyperplane separating the data into the desired classes. 
    
    Computing the parameters of the SVM is achieved by minimizing the expression below, where points in the dataset are represented by \(x_i\), corresponding output classes are \(y_i\), parameters of the classifier are \(w\) and \(b\), and \(\lambda\) is an \(l2\) regularization penalty on the weights. 

\[ \left[ \frac{1}{n} \sum_{i=1}^{n} \max (0, 1 - y_i(w \cdot x_i + b)) \right] + \lambda \left \| w \right \|_2\ \label{eqn:svm}\tag{5}\]

	We constructed a domain specific classifier for each database - computer science and medicine - and a combined classifier for the union of both databases. The classifier was set to output posterior probability distributions for each chosen class using Platt Scaling \cite{svm}.
    
    \subsection{Visual Classifier}
    The visual classifier used a similar bag of words into SVM approach as the text classifier but substituted the weighted term frequency vectors for the clustered SIFT vectors. This time a Gaussian kernel \cite{rbf}, described by equation \eqref{eqn:rbf}, was used for the SVM due to empirically better performance on the image frequency vectors. The Gaussian kernel is implemented by inserting the function described in \eqref{eqn:rbf} into the Lagrangian dual of the minimization in \eqref{eqn:svm}. 
    
\[ k(x_i,x_j) = \exp(- \frac{\left \| x_i-x_j \right \|_2}{2\sigma^2}) \label{eqn:rbf}\tag{6}\]
In the equation above, \(x_i\) and \(x_j\) are the data point and support vector respectively, and \(\sigma\) is a regularizing parameter which affects the smoothness of the decision boundary.  Using a Gaussian kernel allows the SVM to learn a nonlinear function boundary with respect to the input data without explicitly mapping the data into different dimensional space.

As with the text classifiers, domain-specific classifiers were constructed for each database and an overall classifier was constructed by combining both databases.  All classifiers were evaluated using area under the curve (AUC) of the receiver operating characteristic (ROC). An ROC curve represents the relationship between the true positive rate and false positive rate. Using the AUC metric allows us to evaluate our classifier across the entire ROC curve using one integer value.
    
\subsection{Meta Classifier}
 A meta-classifier was constructed to combine the outputs of the text and visual classifiers by running a new linear SVM over their concatenated AUC-scaled output probabilities as shown in in \eqref{eqn:meta1}.
\[m_i = (y_{i}\sqrt{a_{i}})_{c1} \parallel (y_{i}\sqrt{a_{i}})_{c2} \label{eqn:meta1}\tag{7}\] 
The inputs to the SVM were \(m_i\)  the concatenated class confidence \(y_i\) of each classifier \(c\) scaled by the square root of its AUC,  \(\sqrt{a_i}\). The scaling factor was added to up-weight a generally better performing classifier.

 
\subsection{Nonlinear Classifier}
 A nonlinear classifier was also tried alongside the meta classifier. The nonlinear classifier function chose the output from either the text classifier or the image classifier based on an AUC-weighted version of the highest class probability for each function. An additional weight \(\gamma\) was added to the text classifier \(c1\), as it continually outperformed the image classifier \(c2\). The nonlinear classifier algorithm is shown in Equation \eqref{eqn:nonlin}: 

\[ \arg \max_{y_i} ( (y_{i}\sqrt{a_{i}} + \gamma)_{c1}, (y_{i}\sqrt{a_{i}})_{c2}) \label{eqn:nonlin}\tag{8}\] 
where $y_i$ is the  probability that a paper is in a class according to it's corresponding classifier $c$, and $\sqrt{a_i}$ is the square root of the corresponding classifier's AUC. The \(\gamma\) weighting factor was tested for values between 0 and 0.30, and showed a best performance at 0.25. However, the nonlinear classifier never exceeded the performance of the original classifier.

\section{Results}
We evaluated each classifier's performance both on their own domain's database and on the opposite domain. We used AUC as a testing metric for all of our classifiers in order to give a single-number metric for our classifiers' performances at all possible ratios of true positive rate (TPR) to false positive rate (FPR). Table 2 shows the overall results for classifiers predicting on the datasets they were trained on - termed intra-domain classifiers - and Figure 3 shows the overall results for classifiers predicting on the domains they were not trained on - termed inter-domain classifiers. Figure 2 and Figure 3 show the ROC curves for intra-domain classifier performance on the CS and medical domains, respectively.

Within their own domain, the classifiers achieved good accuracy, with AUC's ranging from 0.90 for the medical database to 0.96 for the computer science database. The text classifiers offered stronger accuracy intra-domain, with scores between 0.06 and 0.07 AUC above their image counterparts. The meta classifier further improved intra-domain performance by up to 0.02 AUC over the text classifier. The nonlinear classifier did not improve performance over the text classifier and in many cases lost performance. When using a nonlinear weighting of 0.25 towards the better performing classifier, the nonlinear classifier was never able to match the overall performance of the text classifier. When additional non-linearity was introduced , the classifier simply mirrored the output of the text classifier itself. Although it is shown in Figure 2 and Figure 3 for completeness, the nonlinear classifier is excluded from the results in Table 1 and in discussion due to its poor performance.

When applied on domains they were not trained on, the performance of the classifiers dropped precipitously, as shown in Table 2. AUC of the text classifier ranged from 0.51  to 52, only slightly above random guessing, and the AUC of the image classifier ranged from 0.61 to 62. The meta classifier scores ranged from 0.61 to 62. They did not provide any improvements over the  classifiers they trained on the inter-domain tests and seemed to default to the strongest classifier they were trained on.

When trained and tested on a combined set of all of the papers in both domains, the classifiers did well, scoring 0.93, 0.85 and 0.94, respectively for text, image and meta classifiers as shown in Table 1. The scores on the combined set actually exceeded scores on the medical set, indicating that the discriminating features of the the computer science dataset may be stronger than those of the medical dataset. Again, the text classifier showed better performance than the image classifier, but the meta-classifier showed the best performance of all.
	

\begin{table}[]
	
\begin{tabular}{ |p{1.8cm}|p{1.7cm}||p{1cm}|p{1cm}|p{1cm}|  }
 \hline
 \multicolumn{5}{|c|}{\textbf{Intra-Domain Results}} \\
 \hline
 \multicolumn{2}{|c||}{Classifier Domain}&
 \multicolumn{3}{c|}{Classifier AUC} \\
 \hline
 Training Set & Test Set & Text & Visual & Meta \\
 \hline
 CS & CS  & {0.95} & {0.89} & {0.96}\\
 Medical & Medical& {0.88} & {0.81} & {0.90}\\
 Combined & Combined & {0.93} & {  0.85} & {  0.94}\\
 \hline
\end{tabular}
    \label{fig:intraresults}
	{\\*\\*\\*Table 1. AUC of the various classifiers predicting on the dataset they were trained on. Text classifiers show consistently better results than the visual classifiers, but meta-classifiers perform better than both. Classifiers trained on CS dataset shows the greatest AUC overall.}
\end{table}

\begin{table}[]
	
\begin{tabular}{ |p{1.8cm}|p{1.7cm}||p{1cm}|p{1cm}|p{1cm}|  }
 \hline
 \multicolumn{5}{|c|}{\textbf{Inter-domain Results}} \\
 \hline
 \multicolumn{2}{|c||}{Classifier Domain}&
 \multicolumn{3}{c|}{Classifier AUC} \\
 \hline
 Training Set & Test Set & Text & Visual & Meta \\
 \hline
 CS & Medical& {0.51} & {0.61} &{0.61}\\
 Medical & CS & {0.52} & {  0.62} & {  0.62}\\
 
 \hline
\end{tabular}
    \label{fig:crossresults}
	{\\*\\*\\*Table 2. AUC of the various classifiers predicting outside of the dataset they were trained on. Visual classifiers show consistently better results than the text classifiers. Meta-classifiers do not provide any improvement over the text or visual classifiers in the inter-domain case.}
\end{table}

\begin{figure}[]
	\includegraphics[width=0.5\textwidth]{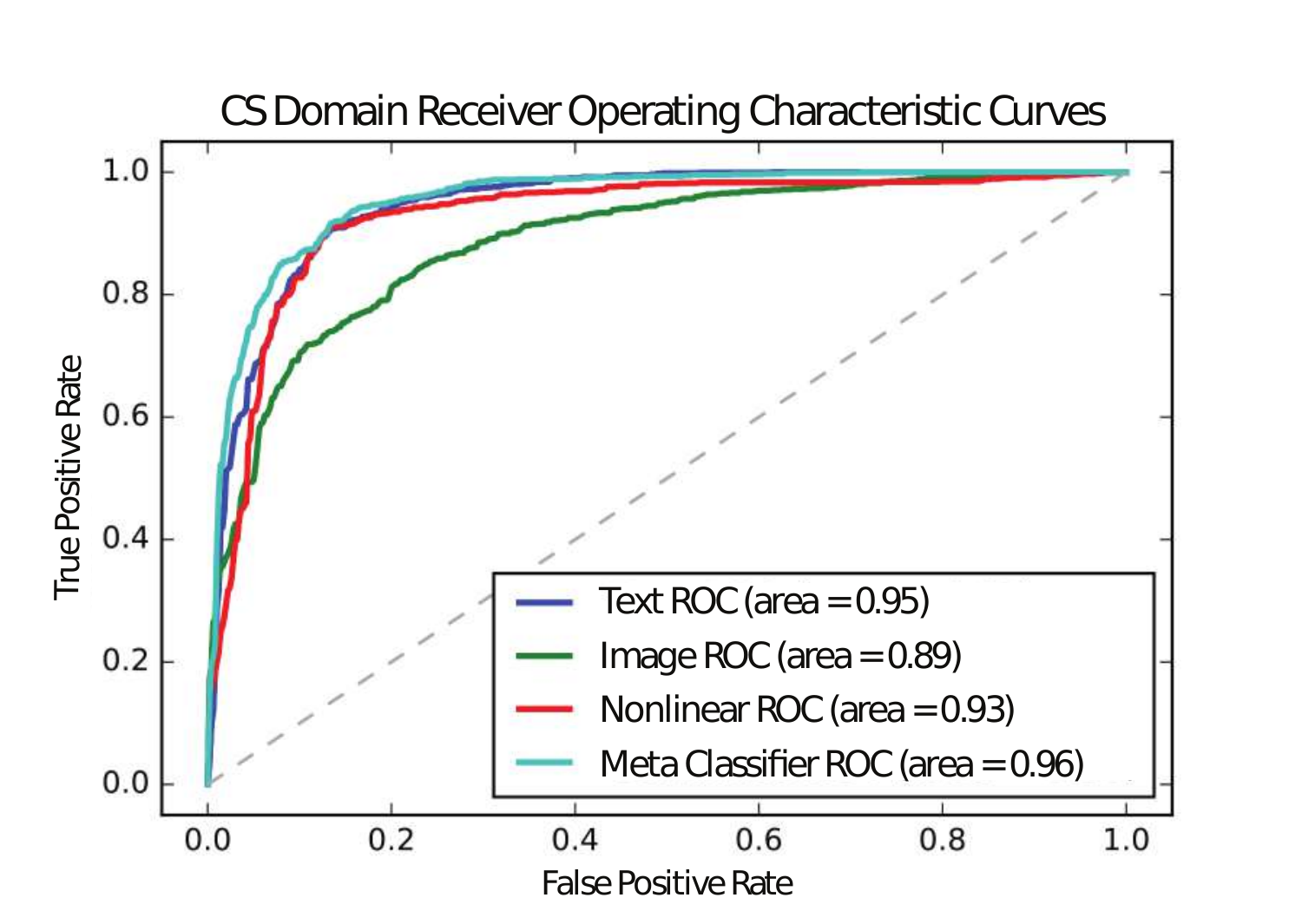}
	\label{fig:csROC}
	\caption{Intra-domain ROC curves for the classifiers on the computer science dataset.}
\end{figure}

\begin{figure}[]
	\includegraphics[width=0.5\textwidth]{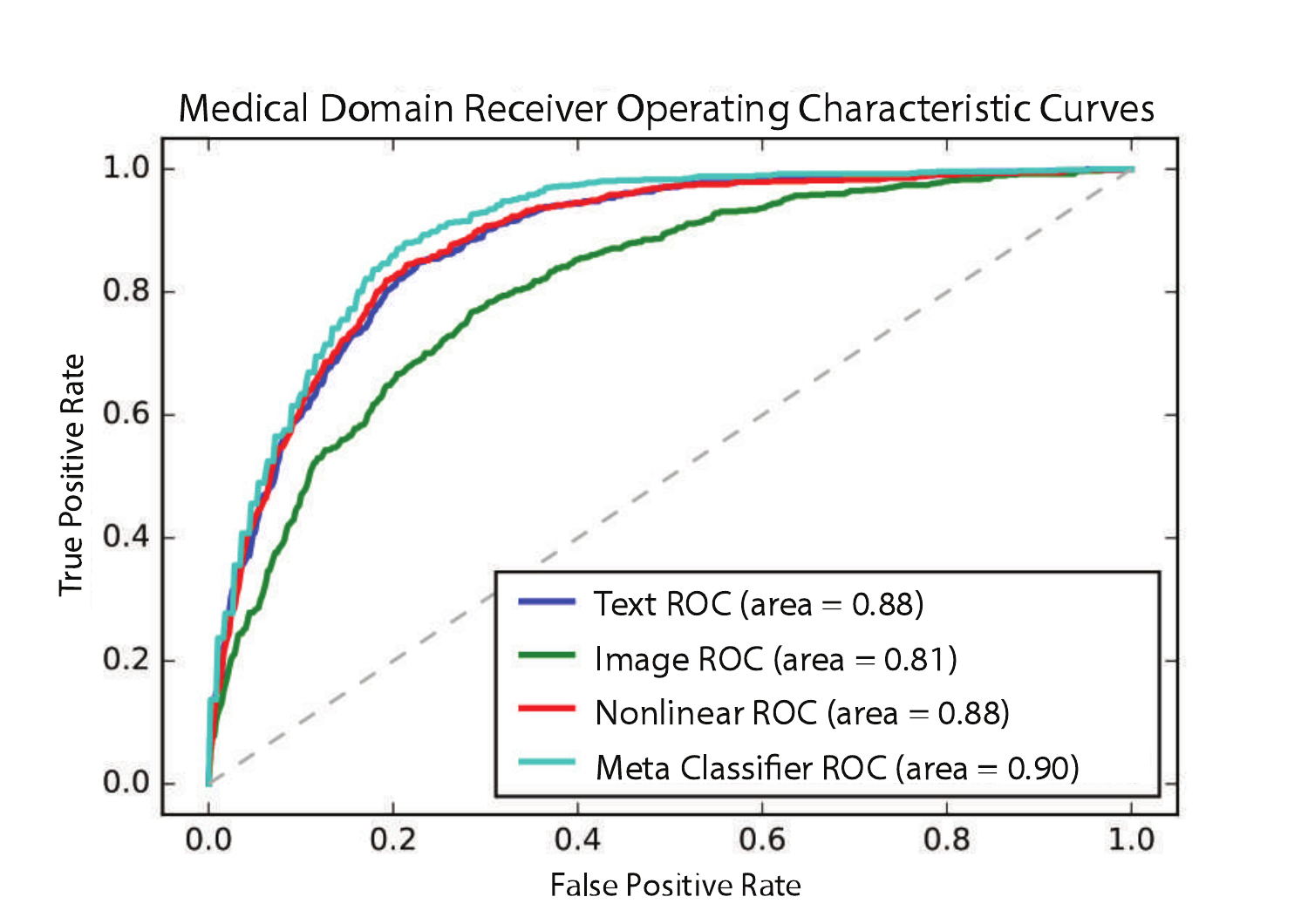}
    \label{fig:medROC}
	\caption{Intra-domain ROC curves for the classifiers on the medical dataset.}
\end{figure}
\section{Analysis \& Discussion}

As expected, both the computer vision and text classifiers performed better on the domain that they were trained on. Although they both performed worse on inter-domain tasks, the performance of the visual classifier fell off less when predicting across domains than the text classifier. The visual classifier lost on average 25\% accuracy when predicting on a the opposite domain whereas the text classifier lost 40\% on average, outputting results no better than random guessing. This increased delta indicates that the text classifiers were more prone to overfitting with respect to domain than the visual classifiers. 

One possible cause for the superior inter-domain performance of the visual classifiers is that k-means clustering of the SIFT features reduced the dimensionality of their inputs. Pegging each SIFT feature to its closest centroid reduces the input space and can be thought of as a form of regularization. Additionally, using a Gaussian kernel for the visual SVM has the equivalent effect of smoothing the input data before classification. Both of these processes may have allowed the classifier to generalize more easily across domains. 

However, a more intuitive explanation lies in the data itself. Upon empirical examination, highly cited papers had relatively consistent gestalts of visually pleasing mixtures of text, figures, tables and equations. Uncited papers tended to have wildly varying gestalts. For example, in Figure 1, the correctly classified uncited paper contains all text with no tables or figures. Other correctly classified uncited documents included large amounts of white-space or overwhelmingly sized full-page figures or tables. It may be expected that highly cited papers would have a similar overall visual composition across domains whereas uncited papers may not. Writers of highly cited papers may be more experienced researchers, with a strong sense of how to convey their ideas in the format of an academic paper and a strong understanding of the culture of their domain and what is expected in terms of gestalt. Conversely, researchers fresher to the field may struggle with effectively translating their research into a cohesive output and may be less familiar with the culture with respect to communicating visually and presentation. 

\begin{table}[]

	\begin{center}
 	\begin{tabular}[]{l|l|l|l|l|}
\multicolumn{0}{c}{\textbf{   Low}}&{\textbf{Citation   }}&
\multicolumn{1}{c}{\textbf{   High}}&{\textbf{Citation   }}\\\hline
\textit{Coeff}&\textit{Feature}&\textit{Coeff}&\textit{Feature}\\\hline
-1.72&quantum&1.51&figure\\\hline
-1.62&phys&1.21&performance\\\hline
-1.18&trada&1.18&graph\\\hline
-1.04&lattice&1.1&user\\\hline
-0.99&et&1.08&web\\\hline
-0.98&queens&1.05&programming\\\hline
-0.94&mpec&1.03&mindfulness\\\hline
-0.9&sas&1.01&acm\\\hline
-0.88&sn&1.0&design\\\hline
-0.87&ipv6&0.99&search\\\hline
-0.86&rfc&0.98&section\\\hline
-0.85&cognitive&0.94&logic\\\hline
-0.85&physics&0.93&theorem\\\hline
-0.82&benes&0.91&use\\\hline
-0.81&field&0.91&gls\\\hline
-0.81&eve&0.89&pages\\\hline
-0.8&x6ti&0.89&sensor\\\hline
-0.8&integration&0.89&service\\\hline
-0.78&arxiv&0.88&nodes\\\hline
-0.78&smoking&0.88&skin\\\hline
	\end{tabular}
    \end{center}
    
    \label{tab:infogain}
\textup{\\*Table 3. The top 20  most informative text features in the combined domains, by term frequency coefficient. Words with low coefficients are most useful for classifying a paper into the uncited class whereas words with high coefficients are most useful for classifying a paper into the highly cited class.}
\end{table}

The text classifiers did very well on the intra-domain task but generalized poorly across domains. This indicates that they may have only been fitting the data based on features specific to the domain itself. To verify this hypothesis, we visualized the top 20 most informative text features by term-frequency coefficient for the combined domains in Table 3. The most discriminating feature for high citations is the unremarkable word ``figure," which likely appears in conjunction with actual figures in a paper. This correlates with our computer vision analysis, confirming that figures are important to our classifiers' learned decision function regardless of our choice of input features.  

The top 20 discriminating terms contain a majority of domain specific words but also many non-domain specific ones. Examples of domain specific terms on the list include ``quantum," ``sas," ``cognitive," ``physics," and non domain-specific words include ``figure," ``pages," ``field," and ``design".  The fact that the most informative words for classification include a mixture of domain-specific and non-domain specific words implies that the text classifiers were fitting on more than just specialized domain knowledge.

 An observation from Table 3 is the juxtaposition of ``acm" in the high citation bin and ``arxiv" in the low citation bin. In this context, ``acm" likely refers to Association of Computing Machinery, one of the worlds largest scientific societies, and ``arxiv" likely refers to arXiv.org, an online archive of pre-prints and e-prints of computer science academic papers. Although arXiv.org contains a large amount of high quality work, surpassing 1 million articles total in 2014 \cite{arxiv}, our classifier has independently learned that papers released in ACM papers have a higher chance of citation than those released on arXiv. In other words our text classifier has learned by itself to judge a paper based on its venue of classification. Useful as it may be to the classifier, the inclusion of publication venue as a strong signal can be viewed as aform of classification bias. In the future it may be useful to filter on publication venue names in order to prevent this type of bias.

Going forward, we would like to test our classifiers on domains which have a culture of using less figures and visualizations than medicine and computer science; perhaps a field withing the humanities. A domain which uses less images may make a visual classifier. Additionally, classifiers trained on domains with strong cultures of visualizations in the published output may suffer when transferred to domains which do not share these cultures and vice-versa.

\section{Conclusions}
	In this paper we set out to design a system that automatically predicts which academic papers will have a high or low citation count based on text and image features. We create our own dataset for the task, one that is larger and broader than previous efforts in the field, spanning entire disciplines of study rather than specific conferences. This allows our conclusions to generalize better relative to previous work. We also use more varied signals than previous work, including text data, visual data and a combination of both. Finally we perform the novel task of comparing the performance of classifiers across domain boundaries.
    
    All  of our classifiers performed well within their own domains with AUC's of up to 0.96. A classifier we trained on a combination of the computer science domain and medical domain also performed well, with an AUC of 0.94, indicating that with more research it may be possible to train a classifier on an domain-agnostic corpus of academic output with favorable results. 
    
    Text classifiers were generally stronger than image classifiers on the intra-domain case and significantly computationally less expensive. By fusing the text and image classifiers together with a meta-classifier we found slightly better performance but not enough to warrant their use on intra-domain classification. 
    
    As expected, performance of all of our classifiers diminished precipitously when classifying a domain the classifier was not trained on. Text classifiers tested on domains they were not trained on fared no better than random guessing. However, we found that classifiers trained on visual features were significantly more robust than their text based counterparts when classifying across domains.

\section*{Acknowledgments}
This work was generously supported in part by Xerox Foundation and New York State through the Goergen Institute for Data Science at the University of Rochester.

\ifCLASSOPTIONcaptionsoff
  \newpage
\fi



%

\pagebreak

%








\end{document}